\documentclass[twocolumn,showpacs,preprintnumbers,amsmath,amssymb]{revtex4}

\usepackage{graphicx}
\usepackage{dcolumn}
\usepackage{bm}

\begin{document}

\preprint{APS/123-QED}

\title{NMR Search for the Spin Nematic State in LaFeAsO Single Crystal}

\author{M. Fu$^{1}$,  D. A. Torchetti$^{1}$, T. Imai$^{1,2}$, F.L. Ning$^{3}$, J.-Q. Yan$^{4,5,6}$, and  A.\ S.\ Sefat$^{4}$}

\affiliation{$^{1}$Department of Physics and Astronomy, McMaster University, Hamilton, Ontario L8S4M1, Canada}
\affiliation{$^{2}$Canadian Institute for Advanced Research, Toronto, Ontario M5G1Z8, Canada}
\affiliation{$^{3}$Department of Physics, Zhejiang University, Hangzhou 310027, China}
\affiliation{$^{4}$Materials Science and Technology Division, Oak Ridge National Laboratory, TN 37831, USA}
\affiliation{$^{5}$Department of Materials Science and Engineering, The University of Tennessee at Knoxville, Knoxville, TN 37996, USA}
\affiliation{$^{6}$Division of Materials Science and Engineering, Ames Laboratory, US-DOE, Ames, Iowa 50011, USA}

\date{\today}


\begin{abstract}
We report a $^{75}$As single crystal NMR investigation of LaFeAsO, the parent phase of a pnictide high $T_c$ superconductor.  We demonstrate that spin dynamics develop a strong two-fold anisotropy within each orthorhombic domain below the tetragonal-orthorhombic structural phase transition at $T_{TO}\thickapprox156$\ K.  This intermediate state with a dynamical breaking of the rotational symmetry freezes progressively into a spin density wave (SDW) below $T_{SDW}\thickapprox142$\ K.  Our findings are consistent with the presence of a spin nematic state below $T_{TO}$ with an incipient magnetic order.
\end{abstract}


\maketitle


The mechanism of high $T_c$ superconductivity in iron pnictides remains enigmatic \cite{Kamihara}.  Earlier NMR measurements demonstrated that low frequency spin fluctuations associated with the spin density wave (SDW) instability grow toward $T_c$ near optimal doping \cite{NakaiJPSJ, NingPRL, GrafePRL, ZhengPRL}, favoring the scenario of spin-fluctuation induced superconductivity.  There is, however, a complication; the slowing of the lattice vibrations accompanies that of spin fluctuations \cite{McGuire, Mandrus, Uchida}.  For example, neutron scattering measurements showed that LaFeAsO undergoes a tetragonal-orthorhombic structural phase transition at $T_{TO}\thickapprox156$\ K, followed by a SDW ordering at $T_{SDW}\thickapprox142$\ K \cite{delaCruz, McGuire, Vaknin}.  Moreover, the softening of the lattice begins at as high as $\sim200$ K, and continues through $T_{TO}$ and $T_{SDW}$ down to $T_{Domains}\thickapprox120$\ K \cite{McGuire}, where the growth of the orthorhombic domains ends \cite{Vaknin, Ricci}.  Theoretical analysis of LaFeAsO  based on the frustrated $J_{1}$-$J_{2}$ model suggests that the Ising symmetry of Fe spins may be already broken below $T_{TO}$ without a three-dimensional magnetic long-range order \cite{Sachdev}.  Moreover, the intermediate temperature range between $T_{TO}$ and $T_{SDW}$ of LaFeAsO may be identified as a {\it spin nematic state} \cite{Kivelson}.  

In such a nematic state, the spin correlations break the tetragonal symmetry, i.e. $\langle{\bf S}_{i} \cdot {\bf S}_{i+x}\rangle = -\langle{\bf S}_{i} \cdot {\bf S}_{i+y}\rangle$ (we refer readers to Fig.\ 1 of a review article \cite{FernandesReview} for a pictorial demonstration of the nematic state).  More recent theoretical analysis based on an itinerant electron picture \cite{Fernandes2} or an orbital fluctuation model \cite{Kontani} also led to analogous conclusions.  The prospect of observing such a magnetic analogue of a liquid crystal below $T_{TO}$ with an incipient (``fluctuating") magnetic order, and its potential link with the mechanism of high $T_c$ superconductivity, has stimulated strong interest among researchers.  The past experimental efforts searching for the signature of nematicity were focused primarily on the BaFe$_{2}$As$_{2}$ series ({\it e.g.} \cite{Analytis, Tanatar, Birgeneau, Nakajima}).  However, the proximity between the structural and SDW transitions, and/or the twinning of orthorhombic domains hampered these efforts.

In this Letter, we report a microscopic $^{75}$As NMR investigation of LaFeAsO for a single crystal \cite{Yan}, and compare our results with neutron scattering \cite{Vaknin} and magnetic susceptibility $\chi$ measured for the same piece of $\sim20$\ mg crystal.  The usage of a single crystal enabled us to resolve complicated changes of NMR lineshapes across $T_{TO}$ and $T_{SDW}$ for the first time, and find the signature of the spontaneous breaking of the rotational symmetry.  We will demonstrate that low frequency spin dynamics indeed exhibit a strong anisotropy within {\it each} orthorhombic domain with a two-fold symmetry below $T_{TO}\thickapprox 156$\ K.  Moreover, the anisotropic spin state freezes progressively into a static SDW from $T_{SDW}\thickapprox 142$\ K to $T_{Domains}\thickapprox120$\ K, and the SDW ordered and paramagnetic domains coexist in a broad range of temperature.   Our findings uncover the presence of an unconventional intermediate spin state below $T_{TO}$ with the signatures of spin nematicity.

In Fig.\ 1, we summarize representative field-swept $^{75}$As NMR lineshapes observed at $f_{NMR} = 58.159$\ MHz (nuclear spin $I=3/2$).    When we apply an external magnetic field $B_{ext}$ along the c-axis, a sharp paramagnetic (PM) {\it central peak} appears at $B_{ext}^{center}\sim7.96$\ T for the $I_{z}=+1/2$ to $-1/2$ transition, as shown in Fig.\ 1(b); the resonant condition is $f_{NMR}=(1 + ^{75}K)\gamma_{n}B_{ext}^{center}$, where the nuclear gyromagnetic ratio $\gamma_{n}/2\pi=7.2919$\ MHz/T.  $^{75}K\sim 0.002\ (\sim 0.2$\ \%) is the Knight shift, which measures the product between the local spin susceptibility and the hyperfine coupling constant.  We note that the full-width at half-maximum of the PM central peak is as sharp as 6.5\ kHz (8.9\ Oe) at 290\ K in the absence of orthorhombic distortion.  The narrow linewidth is comparable to that of undoped BaFe$_2$As$_2$ \cite{Kitagawa, NingJPSJ}, and attests to the high homogeneity of our crystal.  

In Fig.\ 1(a), we observe an $I_{z}=\pm3/2$ to $\pm1/2$ paramagnetic {\it satellite peak} near $B_{ext}^{satellite}=6.7$\ T (connected by a dotted line).  We also found the strongly temperature dependent antiferromagnetic (AF) central peak arising from the statically SDW ordered domains (marked by downward arrows), but only below the onset temperature of $T_{SDW}^{NMR}\thickapprox 135$\ K ($<\ T_{SDW}$).  The AF central peak is shifted from the PM central peak at $B_{ext}^{center}\sim7.96$\ T in Fig.\ 1(b) by a static hyperfine field along the c-axis, $\langle B_{hf}^{c} \rangle$.  $\langle B_{hf}^{c} \rangle$ originates from the ordered Fe magnetic moments $M_{Fe}$, and $\langle B_{hf}^{c} \rangle \propto M_{Fe}$ \cite{Kitagawa}.  In Fig.\ 2(b), we deduce the temperature dependence of $\langle B_{hf}^{c} \rangle$ from Fig.\ 1(a), and compare the results with the neutron scattering data of $M_{Fe}$ \cite{Vaknin}.

\begin{figure}
\centering
\includegraphics[width=3.2in]{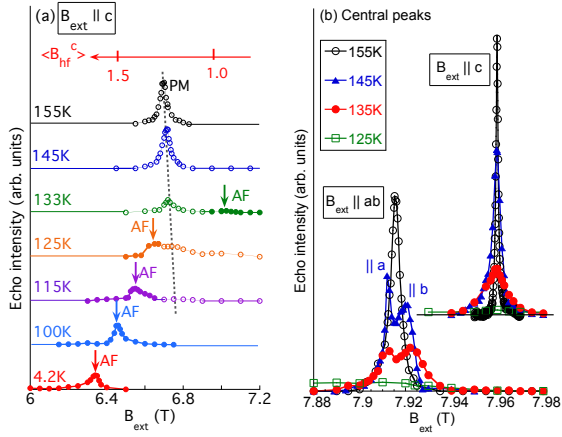}
\caption{\label{Fig1:epsart} (Color online) Representative $^{75}$As NMR lineshapes observed at 58.159\ MHz. (a) The paramagnetic (PM) satellite peak ($\circ$), and the antiferromagnetic (AF) central peak marked with a downward arrow ($\bullet$) in $B_{ext} \parallel c$.  For clarity, the vertical axis is shifted for different temperatures. The dotted line indicates the shift of the PM satellite peak with temperature.  The horizontal axis at the top of (a) measures the split $\langle B_{hf}^{c} \rangle$ from the PM central peak at 7.96\ T in (b).  (b) The PM central peak with $B_{ext} \parallel c$ (upper right) and $B_{ext} \parallel ab$ (lower left).  For clarity, the origin of the vertical axis is shifted for $B_{ext} \parallel c$.  The formation of twinned orthorhombic domains  along the two orthogonal axes, as schematically shown in Fig.\ 2(a), results in the splitting of the $B_{ext} \parallel ab$ lineshape  below $T_{TO}\thickapprox156$\ K.
}
\end{figure}

The splitting between the central and satellite peaks, $\nu_Q^{c} = \gamma_{n} (B_{ext}^{center} - B_{ext}^{satellite}) \thickapprox9.2$\ MHz, measures the nuclear quadrupole interaction with the electric field gradient (EFG).  We summarize the temperature dependence of $\nu_Q^{c}$ in Fig.\ 2(c).  The EFG is the second derivative of the Coulomb potential arising from electrons and ions near the observed $^{75}$As sites.  Below $T_{TO}\thickapprox156$\ K, $\nu_Q^{c}$ \ exhibits a sharp downturn, because the EFG is sensitive to the local structural environment.   Earlier diffraction measurements showed that the growth of orthorhombic domains finally comes to an end at $T_{Domains}\thickapprox 120$\ K \cite{Vaknin, Ricci}; $\nu_Q^{c}$ also levels off below $T_{Domains}$.  

\begin{figure}
\centering
\includegraphics[width=3.2in]{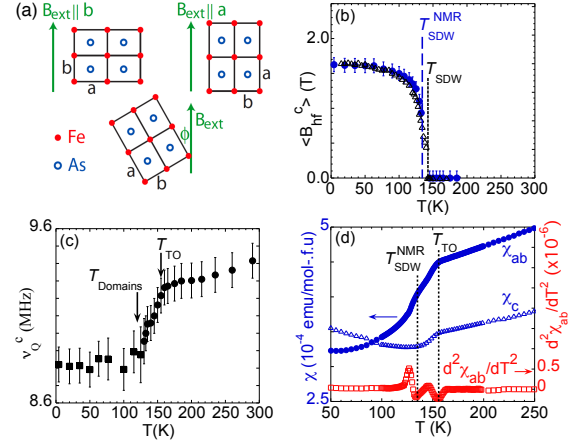}
\caption{\label{Fig2:epsart} (Color online) (a) Schematics of the orthorhombic domain with $B_{ext} \parallel b$ (upper left) and $B_{ext} \parallel a$ (upper right) within twinned ab-planes.  Bottom: rotation by azimuthal angle $\phi$.  (b) ($\bullet$): $\langle B_{hf}^{c} \rangle$ deduced from Fig.\ 1(a).  ($\triangle$): the sub-lattice magnetization $M_{Fe}$ deduced as the square-root of the magnetic Bragg scattering intensity (after ref.\cite{Vaknin}), normalized at 4.2\ K.  (c) $\nu_{Q}^{c}$ measured for the paramagnetic ($\bullet$) and antiferromagnetic  ($\blacksquare$) peaks.  (d) The magnetic susceptibility $\chi_{ab}$ ($\bullet$) and $\chi_{c}$ ($\triangle$) measured in 1\ T with dc-SQUID.  Also shown is the second derivative of $\chi_{ab}$ ($\square$).  Notice the presence of two kinks in $\chi_{ab}$ at $\sim156$\ K and $\sim133$\ K, as evidenced by the negative maxima of $d^{2}\chi_{ab}/dT^{2}$.       
}
\end{figure} 

In Fig.\ 1(b), we also show representative NMR lineshapes with $B_{ext}$ applied along the tetragonal a-axis within the FeAs planes.  Above $T_{TO}$, all tetragonal domains are equivalent, hence we observe a single central peak.  Below $T_{TO}$, the elongated a-axis of each domain points either along the direction of $B_{ext}$, or orthogonal to $B_{ext}$, due to twinning of the orthorhombic domains, as sketched in Fig.\ 2(a).  Moreover, orthorhombic distortion breaks the axial symmetry of the EFG.  The difference in the second order effects of the nuclear quadrupole interaction thus results in splitting of the central peak; the peak positions depend on the direction of the elongated a-axis relative to $B_{ext}$.  We confirmed that the NMR line splitting in Fig.\ 1(b) is caused entirely by the difference in the second order nuclear quadrupole effects, which is inversely proportional to $B_{ext}$ \cite{Takigawa}.  That is, the NMR Knight shift is still axially symmetric, and shows very little temperature dependence  below $T_{TO}$, $^{75}K_{a}\ =\ ^{75}K_{b} \simeq 0.22\pm0.03$ \%.   In view of the fact that expansion of the lattice from $T_{TO}$ to 290\ K results in a larger value of the quadrupole frequency as shown in Fig.\ 2(c), we tentatively assign the peak near $B_{ext}\sim7.91$\ T to the orthorhombic domains with $B_{ext}\parallel a$; the other peak near $B_{ext}\sim7.92$\ T arises from the orthorhombic domains with $B_{ext}\parallel b$ instead.  From the splitting of an $I_{z}=\pm3/2$ to $\pm1/2$ satellite peak, we estimate $\nu_Q^{a}\sim 5.2$\ MHz and $\nu_Q^{b}\sim 3.9$\ MHz at 145\ K.  We also confirmed that these double peaks collapse into one when we apply $B_{ext}$ along the [110] direction within the ab-plane, because $B_{ext}$ points along the diagonal direction for all orthorhombic domains in such a geometry.  

For bulk averaged measurement techniques such as resistivity, the twinning of orthorhombic domains would result in experimental data averaged over two orthogonal directions within the ab-plane, unless one applies a uniaxial stress  \cite{Analytis, Tanatar, Birgeneau, Nakajima}.  It is not straightforward, then, to probe the {\it spontaneous} breaking of the rotational symmetry.  In contrast, NMR is a local probe.  Since we have succeeded in resolving the central peaks in Fig.\ 1(b) for different orientations, we can investigate the in-plane anisotropy of spins within each orthorhombic domain.

\begin{figure}
\centering
\includegraphics[width=3.2in]{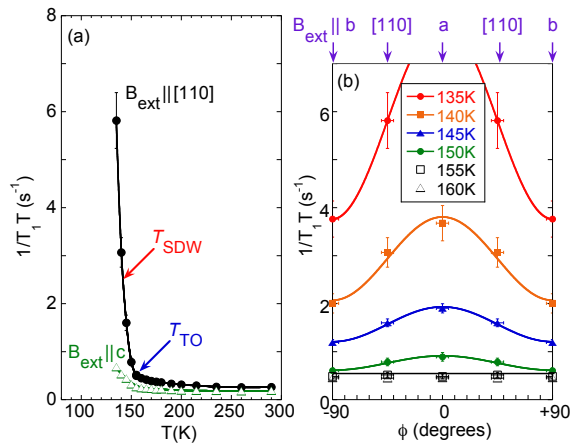}
\caption{\label{Fig3:epsart} (Color online) (a) Temperature dependence of $1/T_{1}T$ measured for the PM central peak with $B_{ext}$ along the [110]  ($\bullet$), and the c-axis ($\triangle$) direction. The solid curves are guides to the eye only.  (b) $1/T_{1}T$ as a function of the field orientation $\phi$ as defined in Fig.\ 2(a), within each orthorhombic domain.  The results are symmetrized for $\phi \leq 0$ due to crystal symmetry.  Below $T_{TO}\sim156$\ K, $1/T_{1}T$ exhibits increasingly strong dependence on the in-plane orientation of $B_{ext}$, which can be modeled by the function $1/T_{1}T=A+B cos^{2}\phi$ (solid curves).  The accidental superposition between different NMR peaks makes accurate measurements of $1/T_{1}T$ unfeasible below 135-140\ K, depending on the direction of $B_{ext}$.
}
\end{figure}

In Fig.\ 3(a), we summarize the temperature dependence of the nuclear spin-lattice relaxation rate $1/T_{1}$ divided by $T$, $1/T_{1}T$, for the PM central peaks \cite{T1fit}.   Our preliminary $1/T_{1}T$ results for the SDW ordered state are very similar to the case of BaFe$_{2}$As$_{2}$ \cite{Kitagawa}, and beyond the scope of the present work.  $1/T_{1}T$ measures low-frequency spin fluctuations: $(1/T_{1}T)_{\alpha} \propto \Sigma_{{\bf q}}|F({\bf q})|^{2}\chi"({\bf q},f_{NMR})/f_{NMR}$, where ${\alpha}$ specifies the direction of $B_{ext}$, $|F({\bf q})|^{2}$ and $\chi"$ are the hyperfine form factor \cite{Shannon} and the imaginary part of the dynamical electron spin susceptibility, respectively, and the {\bf q} summation is taken within the Brillouin zone.  $1/T_{1}T$ shows a mild increase with decreasing temperature  down to $T_{TO}$ due to the slow growth of a short range SDW order.  Once we enter the orthorhombic phase below $T_{TO}$, $1/T_{1}T$ measured with $B_{ext} \parallel$ [110], a-, and b-axis begins to show rapid growth.  This implies that the orthorhombic distortion enhances low frequency components of antiferromagnetic spin fluctuations, and that the dynamic SDW is rapidly slowing down \cite{NingPRL, NingJPSJ, NakaiPRB}.  This conclusion is consistent with the downturn of $\chi$ below $T_{TO}$ \cite{McGuire}, also observed for our crystal as shown in Fig.\ 2(d).     At first glance, the enhancement of $1/T_{1}T$ below $T_{TO}$ is much weaker with $B_{ext} \parallel $ c.  This is simply because the transferred hyperfine fields at $^{75}$As sites from their four nearest-neighbor Fe sites are geometrically cancelled out within the ab-plane in this configuration, i.e. $|F({\bf Q})|^{2}=0$  for the SDW ordering wave vectors ${\bf Q}$, and the contributions of AF spin fluctuations to $(1/T_{1}T)_{c}$ are ``filtered out" \cite{Shannon}.   The growing anisotropy of $1/T_{1}T$ between the ab- and c-axis orientations observed below $T_{TO}$ therefore has little to do with that of the critical dynamics of Fe spins near $T_{SDW}$.

Next, let's turn attention to the angular dependence of $1/T_{1}T$ within the ab-plane, which has been proposed as a novel probe of spin nematicity \cite{Shannon}.  As summarized in Fig.\ 3(b), we don't observe any $\phi$-dependence of $1/T_{1}T$ above $T_{TO}$.  Once we enter the orthorhombic phase below $T_{TO}$, $1/T_{1}T$ begins to develop a strong anisotropy within each orthorhombic domain.  The anisotropy reaches  as much as a factor of $\sim 2$ by $\lesssim 140$\ K.  In view of the very small difference between the lattice constants of the a- and b-axis ($\sim 0.5$ \%) \cite{delaCruz,Yan, Vaknin}, our finding is quite unexpected for paramagnetic spin fluctuations.

As explained above, the Knight shift remains axially symmetric within experimental uncertainties below $T_{TO}$.  It is therefore unlikely that the uniform spin susceptibility or the hyperfine form factor $|F({\bf q})|^{2}$ develops a sizable anisotropy within the ab-plane below $T_{TO}$.   We therefore conclude that low frequency Fe spin dynamics as reflected in $\chi"({\bf q},f_{NMR})$ locally develop a strong rotational anisotropy by a factor of $\sim 2$ within {\it each} orthorhombic domain below $T_{TO}$, without exhibiting a three-dimensional magnetic order.   We note that the intensities of the a- and b-peaks in Fig.\ 1(b) are comparable, hence FeAs planes are randomly twinned.  When averaged over the entire single crystal, Fe spin dynamics would appear almost isotropic within the ab-plane.    

\begin{figure}
\centering
\includegraphics[width=3.2in]{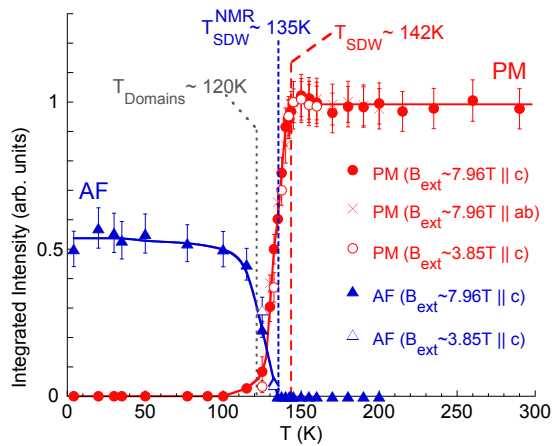}
\caption{\label{Fig4:epsart} (Color online) The temperature dependence of the PM and static AF volume fractions at the NMR time scale, measured as the integrated intensity of the PM and AF central peaks (normalized by the Boltzmann factor).  The solid curves are guides to the eye.  Notice that the PM phase coexists with the AF phase even below $T_{SDW}$ down to $T_{Domains}$.  We measured the spin echo with a delay time $2\tau =40\ \mu$s; the transverse relaxation time $T_{2}$ for the {\it observable} NMR signals is always long ($\sim$msec), and does not appreciably affect the intensity.  Notice that lowering $B_{ext}$ from $\sim 7.96$\ T to $\sim3.85$\ T does not alter the qualitative features.  
}
\end{figure}

The anomalous behavior of Fe spins below $T_{TO}$ is not limited to the in-plane anisotropy of their dynamics.   In Fig.\ 4, we summarize the temperature dependence of the integrated intensities of the NMR signal.  An unusual aspect of the signal intensity is that the paramagnetic NMR peaks don't disappear suddenly at $T_{SDW}\thickapprox 142$\ K, and linger well below $T_{SDW}$ down to $T_{Domains}\thickapprox 120$\ K.  On the other hand, AF NMR signals from statically SDW ordered domains emerge progressively, only below $T_{SDW}^{NMR}\thickapprox 135$\ K ($< T_{SDW}$).  Notice that the PM central peak is still clearly observable below $T_{SDW}$ in Fig.\ 1(b).  Our finding that the PM signal intensity is as large as $\sim 60$\ \%  at $T_{SDW}^{NMR}$ implies that $\sim 60$\ \% of the sample volume remains paramagnetic at 135\ K.  That is, AF and PM domains coexist even below $T_{SDW}^{NMR}$, although neutron scattering begins to detect magnetic Bragg peaks below $T_{SDW}$ \cite{Vaknin}.  In addition, $1/T_{1}T$ measured for the residual PM peak does not blow up at $T_{SDW}$, and continues to increase.  If the SDW ordering in our LaFeAsO single crystal was a typical second order phase transition,  $1/T_{1}T$ would diverge at $T_{SDW}$ due to the critical slowing down of spin fluctuations, followed by a sudden disappearance of PM NMR signals below $T_{SDW}$.  

The reason NMR and neutron scattering detect different onset temperatures of the SDW is that each experimental probe has a different characteristic measurement time scale \cite{Hunt, Mazin}.  Elastic neutron scattering measurements would consider the SDW `static' when fluctuations slow down to below the instrument resolution of $\sim 1$meV.  This means that neutron scattering can take an instantaneous picture of Fe spins with a  `shutter speed' of $\sim10^{-11}$\ sec even if they are still slowly fluctuating.  In contrast, AF NMR signals in Fig.\ 1(a) become observable only when Fe spins in the SDW become static, to the extent that $\langle B_{hf}^{c} \rangle$ is time-independent over the duration of our spin echo measurements, $40\ \mu$sec.  If Fe moments are fluctuating faster than $40\ \mu$sec in some segments of FeAs planes, NMR would see them as motionally averaged out (i.e. paramagnetic).  In other words, the different onset temperatures of the SDW ordering between neutron and NMR data indicate that the fluctuations of the SDW continue to slow down from $T_{SDW}\thickapprox142$\ K to $T_{SDW}^{NMR}\thickapprox135$\ K.  The coexistence of PM and AF domains below 142\ K implies that the fluctuation time scales have a broad distribution throughout the FeAs planes.  We also found that $\chi_{ab}$ exhibits an additional kink at $\sim 133$\ K, as shown in Fig.\ 2(d).  Since SQUID measures the time-independent response of spins, the acceleration in the suppression of $\chi_{ab}$ below  $T_{SDW}^{NMR}$ is consistent with NMR. 

It is interesting to realize that these unusual behaviors of LaFeAsO share similarities with the spin stripes in high $T_{c}$ cuprates \cite{Tranquada}, where the concept of namaticity was originally proposed for unconventional superconductors \cite{Fradkin}.  In the striped cuprates, the spin stripes progressively slow down below a charge ordering at $\sim70$\ K \cite{Tranquada} (instead of $T_{TO}$); elastic neutron scattering, $\mu$SR, and NMR detect the emergence of a static SDW at their respective measurement time scale below $\sim50$\ K \cite{Tranquada}, $\sim30$\ K \cite{Nachumi}, and $\sim1.6$\ K \cite{Hunt}, respectively.  Moreover, PM NMR signals linger well below $\sim 50$\ K \cite{Hunt}. 

The temperature dependence of the AF NMR signal intensity in Fig.\ 4 indicates that the volume fraction of the static SDW at the time scale of NMR gradually increases from $T_{SDW}^{NMR}\thickapprox135$\ K toward $T_{Domains}\thickapprox 120$\ K.  Once the growth of the orthorhombic domains ends at $T_{Domains}$ \cite{Vaknin, Ricci}, the intensity saturates at $\sim 60$\% of the paramagnetic intensity above $T_{SDW}$.  The missing signal intensity suggests that $\langle B_{hf}^{c} \rangle$ is still modulating and/or $T_{2}$ is too fast for a spin echo NMR signal to form in $\sim 40$\ \% of the sample volume.  Our finding is consistent with the earlier $\mu$SR measurements for a powder sample of LaFeAsO; the muon precession with a well-defined frequency of $\sim23$ MHz takes place only in $\sim 70$\ \% of the sample volume, and $\sim 30$\ \% of the $\mu$SR signal is strongly damped \cite{Klauss}.  It remains to be seen whether these ultra slow dynamics of the SDW in 30-40\ \% of the sample volume are caused by the motion of the anti-phase domain boundaries \cite{Mazin} and/or the finite size effects of the orthorhombic domains.  

To summarize, we demonstrated that Fe spin fluctuations in a LaFeAsO single crystal begin to slow down below $T_{TO}\thickapprox156$\ K, accompanied by the local breakdown of the rotational symmetry of spin fluctuations in each of the randomly twinned orthorhombic domains.  The way that the static SDW develops is also unconventional.  The paramagnetic and SDW ordered domains coexist in a wide range of temperature below $T_{SDW} \thickapprox142$\ K due to a distribution in the fluctuation time scales of the SDW.  A large volume of FeAs planes sees freezing of the static SDW only below $T_{Domains} \thickapprox120$\ K.   Our findings point towards the presence of a novel spin state between $T_{TO}$ and $T_{Domains}$ with a dynamically broken rotational symmetry and an incipient magnetic order, i.e. the elusive {\it spin nematic sate}. 

The work at McMaster was supported by NSERC and CIFAR.  The work at Zhejiang was supported by National Basic Research Program of China (No.2011CBA00103) and NSF of China (No. 11274268).  Research at ORNL was supported by the Department of Energy, Basic Energy Sciences, Materials Sciences and Engineering Division.  TI acknowledges helpful discussions with K.\ Ishida, H. Eisaki, S. Uchida, I.\ Mazin, J.\ Tranquada, B.\ Buechner, P.\ Canfield, and B.\ Gaulin.  JQY thanks B.\ Jensen, K.\ Dennis, R.\ McCallum, and T.\ Lograsso for
their assistance in crystal growth, which was also supported by the U.S. DOE.\\



\end{document}